\documentclass{article}

\usepackage{amsmath}
\usepackage{amssymb}
\usepackage{amsfonts}
\usepackage{bm}
\usepackage[dvips,xdvi]{graphicx}

\newcommand{\tr}{\operatorname{Tr}}

\newlength{\refwidth}
\begin{document}


\title{\bf Radius of Gyration in Shear Gradient Direction Governs
Steady Shear Viscosity of Rouse-Type Model}
\author{Takashi Uneyama \\
Department of Materials Physics, Graduate School
 of Engineering, 
Nagoya University, \\
Furo-cho, Chikusa, Nagoya 464-8603, Japan}


\maketitle

\begin{abstract}
We analyze the stress tensor and the gyration tensor of an unentangled
polymer melt under flow by using a Rouse-type single chain model.
We employ the bead-spring type single chain model, in which
beads interact each other via nonlinear potentials such as
the finite-extensible nonlinear elasticity (FENE) potential.
Beads are assumed to obey the Langevin equation with a constant friction coefficient.
We derive simple yet general relations between the stress tensor
and the gyration tensor for this Rouse-type model,
without any additional approximations.
Various formulae for rheological quantities in terms of the gyration tensor
can be derived from the general relations.
For example, the steady shear viscosity is governed by the
gyration radius in the shear gradient direction.
\end{abstract}


%

\section{INTRODUCTION}
\label{introduction}

Linear viscoelasticity of an unentangled polymer melt can be well describe
by the Rouse model\cite{Doi-Edwards-book,Rouse-1953}. The
Rouse model is a single chain dynamics model originally developed for a
chain in a dilute solution, but nowadays it is interpreted as the dynamics model of a
single tagged chain in an unentangled polymer melt.
In the Rouse model, a single tagged chain is modeled as an ideal bead-spring chain.
Namely, the beads without excluded volume interactions are connected by harmonic
springs to form a polymer chain.
The beads are assumed to obey the overdamped Langevin equation
with a constant friction coefficient.
Due to its simple form, 
we can analytically calculate various rheological quantities such as the relaxation
modulus.

One drawback of the Rouse model is that it cannot describe some
nonlinear rheological properties. The Rouse model predicts that the
shear viscosity is independent of the applied shear rate (in the same way
as the harmonic dumbbell model).
Experimental data show that unentangled polymer melts exhibit the shear thinning behavior
under fast shear flows\cite{Stratton-1972,Colby-Boris-Krause-Dou-2007,Matsumiya-Sato-Chen-Watanabe-2022}.
To reproduce the shear thinning behavior, we need to modify the original
Rouse model. For example, by replacing the harmonic spring potential by a nonlinear
bond potential such as the finite-extensible nonlinear elasticity (FENE) potential,
the shear thinning behavior can be reproduced.
However, such modification makes the model complicated and 
not analytically tractable. Sometimes we need approximations such
as the pre-averaging approximation\cite{Watanabe-Matsumiya-Sato-2021}.

In this work, we study the rheological properties of the Rouse-type model
with given nonlinear interaction potentials.
We derive simple yet general relations which holds between the stress tensor
and the gyration tensor, without any additional approximations.
The general relations can be used to derive formulae for some nonlinear rheological
quantities such as the steady shear viscosity in terms of the gyration tensor.
For example, the steady shear viscosity is shown to be governed by the
gyration radius in the shear gradient direction.
We discuss properties of our formula by considering the zero strain rate limit.
We compare our theoretical formula with the viscosity of the Rouse model
and Molecular dynamics (MD) simulation data in the literature.
A possible generalization to systems with friction reduction effect
is also discussed.

\section{MODEL}
\label{model}

\begin{figure}[tb]
\begin{center}
 \includegraphics[width=1.0\refwidth]{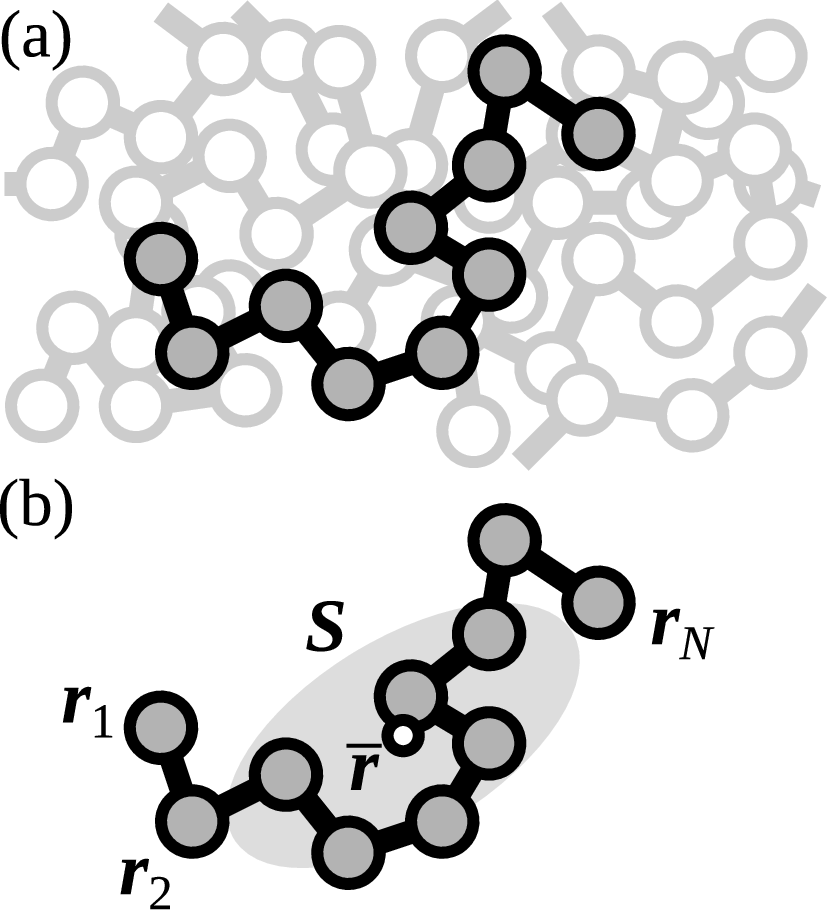}%
\end{center}
\caption{(a) An unentangled polymer melt which consists of many chains.
 A tagged chain (shown by gray circles) is interacting with
 the surrounding chains (shown by white circles).
 (b) The Rouse-type model considered in this work. The dynamics of one
 tagged chain in an unentangled polymer melt is approximately expressed
 by the Langevin equation. The position of the $i$-th bead is expressed as
 $\bm{r}_{i}$, and the center of mass position is expressed as $\bar{\bm{r}}$
 (the white dot).
 The gray ellipse represents the gyration tensor $\bm{S}$.
 \label{rouse_type_model_image}}
\end{figure}

We consider an unentangled polymer melt which consists of $M$ polymer
chains. One polymer chain consists of $N$ beads.
We employ a single chain dynamics model, and
assume that the dynamic equations for individual chains
are not coupled. Then we need to analyze only one tagged chain in the system.
Figure~\ref{rouse_type_model_image} shows images of an unentangled polymer
melt and a single tagged chain.
In the Rouse model, an ideal chain which consists of harmonic springs
is used\cite{Doi-Edwards-book}. In this work, we consider a more general bead-spring type model.
Beads can interact each other via
nonlinear bond potentials (such as the FENE potential)
and also via some interaction potentials (such as
the Lennard-Jones (LJ) potential).
We describe the position of the $i$-th bead in the tagged chain as $\bm{r}_{i}$
($i = 1,2,\dots,N$), and
consider that the interaction potential for a single polymer chain
is given as
\begin{equation}
 \label{interaction_potential}
 \mathcal{U}(\lbrace \bm{r}_{i} \rbrace) = \sum_{i > j} u_{ij}(|\bm{r}_{i} - \bm{r}_{j}|).
\end{equation}
Here, $u_{ij}(r)$ is the interaction potential between the $i$-th and $j$-th
beads. We do not require the beads to be connected linearly.
As long as $N$ beads form a single molecule and they cannot be separated
at the long time limit, any connections are allowed.
For example, we can handle chains with branches and loops.
Also, the interaction potential $u_{ij}(r)$ is not necessarily to
be short-ranged. Coulomb and screened Coulomb potentials, which are important for
polyelectrolytes, can be employed.

We assume that the beads obey the overdamped Langevin equation
in the same way as the Rouse model:
\begin{equation}
 \label{langevin_equation}
 \frac{d\bm{r}_{i}(t)}{dt} = - \frac{1}{\zeta} \frac{\partial  \mathcal{U}(\lbrace \bm{r}_{i}(t) \rbrace)}{\partial \bm{r}_{i}(t)}
  + \bm{\kappa} \cdot \bm{r}_{i}(t)
  + \sqrt{\frac{2 k_{B} T}{\zeta}} \bm{w}_{i}(t),
\end{equation}
where $\zeta$ is the friction coefficient, $\bm{\kappa}$ is the
velocity gradient tensor,
$k_{B}$ is the Boltzmann constant, $T$ is the temperature, and
$\bm{w}_{i}(t)$ is the Gaussian white noise.
Both $\zeta$ and $\bm{\kappa}$ are assumed to be constant.
$\bm{w}_{i}(t)$ satisfies the
following fluctuation-dissipation relation:
\begin{equation}
 \langle \bm{w}_{i}(t) \rangle = 0, \qquad
 \langle \bm{w}_{i}(t) \bm{w}_{j}(t') \rangle = \delta_{ij} \delta(t - t') \bm{1}.
\end{equation}
$\langle \dots \rangle$ represents the statistical average and
$\bm{1}$ is the unit tensor. 
We interpret eq~\eqref{langevin_equation} as the Ito-type
stochastic differential equation\cite{Gardiner-book}.
(The stochastic term in eq~\eqref{langevin_equation} is so-called the
additive noise, and thus the result is not changed if
we interpreted eq~\eqref{langevin_equation} as the Stratonovich-type
Langevin equation.)
Under fast flows, eq~\eqref{langevin_equation} may need some modifications.
We will discuss possible modifications in Sec.~\ref{discussions}.

The single chain stress tensor can be expressed as
\begin{equation}
 \label{single_chain_stress_tensor}
 \hat{\bm{\sigma}}_{\text{single}}(\lbrace \bm{r}_{j} \rbrace)
  = \frac{1}{V} 
  \left[ \sum_{i = 1}^{N} \frac{\partial  \mathcal{U}(\lbrace \bm{r}_{i} \rbrace)}{\partial \bm{r}_{i}} \bm{r}_{i}
  - N k_{B} T \bm{1} \right],
\end{equation}
where $V$ is the volume of the system.
The stress tensor of the system
can be expressed by using the statistical average of
eq~\eqref{single_chain_stress_tensor}:
\begin{equation}
 \label{stress_tensor}
 \bm{\sigma} = M \langle \hat{\bm{\sigma}}_{\text{single}} \rangle
  = \frac{\rho}{N}
  \left[ \sum_{i = 1}^{N} \frac{\partial  \mathcal{U}(\lbrace \bm{r}_{i} \rbrace)}{\partial \bm{r}_{i}} \bm{r}_{i}
  - N k_{B} T \bm{1} \right].
\end{equation}
Here, $\rho = M N / V$ is the bead density.
The stress tensor is symmetric: $\bm{\sigma}^{\mathrm{T}} = \bm{\sigma}$
(The superscript $\mathrm{T}$ represents the transpose).
In equilibrium, the
stress tensor reduces to a simple isotropic tensor which expresses
the pressure of an ideal gas with the number density $\rho / N$:
\begin{equation}
 \label{equilibrium_stress_tensor}
  \bm{\sigma}_{\text{eq}} = - \frac{\rho}{N} k_{B} T \bm{1}.
\end{equation}

To study the average conformation, we define the gyration tensor:
\begin{equation}
 \label{gyration_tensor}
 \bm{S} = \frac{1}{N} \sum_{i = 1}^{N} \langle (\bm{r}_{i} - \bar{\bm{r}}) (\bm{r}_{i} - \bar{\bm{r}}) \rangle ,
\end{equation}
with the center of mass position defined as
\begin{equation}
 \label{center_of_mass_position}
 \bar{\bm{r}} = \frac{1}{N} \sum_{i = 1}^{N} \bm{r}_{i}.
\end{equation}
Intuitively, eq~\eqref{gyration_tensor} represents the characteristic
chain dimension. (See Figure~\ref{rouse_type_model_image}(b).)
The gyration radius in the $\alpha$-direction ($\alpha = x, y,$ and $z$)
becomes $R_{g,\alpha} = \sqrt{S_{\alpha\alpha}}$ and the gyration radius
is given as $R_{g} = \sqrt{\tr \bm{S}}$.

\section{ANALYSIS}
\label{analysis}

Due to the nonlinearlity
of the Rouse-type model, we cannot utilize the Rouse modes\cite{Doi-Edwards-book} for our analysis.
In this section, we introduce the relative bead positions to center of
mass position, in order to simplify the expressions.
Then we show that simple yet general relations hold between the stress tensor $\bm{\sigma}$
and the gyration tensor $\bm{S}$.

\subsection{Relative Positions to Center of Mass}

First we decompose the dynamics into the those for the center of mass position
and other degrees of freedom.
The potential force acting on the center of mass cancels because it
is the sum of internal forces.
From eqs~\eqref{langevin_equation} and 
\eqref{center_of_mass_position}, we have
\begin{equation}
 \frac{d\bar{\bm{r}}(t)}{dt} = \bm{\kappa} \cdot \bar{\bm{r}}(t) + \sqrt{\frac{2 k_{B} T}{\zeta}} \bar{\bm{w}}(t),
\end{equation}
with
\begin{equation}
 \label{langevin_equation_cm}
  \bar{\bm{w}}(t) = \frac{1}{N} \sum_{i = 1}^{N} \bm{w}_{i}(t).
\end{equation}
$\bar{\bm{w}}(t)$ is a sum of Guassian white noises and thus it also
becomes a Guassian white noise. The first and second order moments are
\begin{equation}
 \langle \bar{\bm{w}}(t) \rangle = 0, \qquad
 \langle \bar{\bm{w}}(t)\bar{\bm{w}}(t') \rangle = \frac{1}{N}\bm{1}\delta(t - t').
\end{equation}
Thus we find that the center of mass exhibits a simple Brownian motion
under flow, with the friction coefficient $\bar{\zeta} = \zeta N$.
(This is the same as the behavior of the zeroth Rouse mode in the
Rouse model.)
The stress tensor by the center of mass becomes that of an ideal gas:
\begin{equation}
 \label{stress_tensor_cm}
 \bar{\bm{\sigma}} = - \frac{M}{V} k_{B} T \bm{1}
  = - \frac{\rho}{N} k_{B} T \bm{1}.
\end{equation}
Eq~\eqref{stress_tensor_cm} is independent of time, and is identical
to the equilibrium stress tensor \eqref{equilibrium_stress_tensor}: $\bar{\bm{\sigma}} = \bm{\sigma}_{\text{eq}}$.

Next we introduce the relative positions to express the internal degrees of freedom.
We define the relative position of the $i$-th bead to the center of mass position as
\begin{equation}
 \label{relative_position}
 \bm{q}_{i} = \bm{r}_{i} - \bar{\bm{r}}.
\end{equation}
Here, we note that the relative positions are not linearly independent: $\sum_{i = 1}^{N} \bm{q}_{i} = 0$.
From the transnational symmetry,
the interaction potential depends only on the relative positions: $\mathcal{U}(\lbrace \bm{r}_{i} \rbrace) = \mathcal{U}(\lbrace \bm{q}_{i} \rbrace)$.
Also, the force acting on a bead is unchanged under the variable transform from $\lbrace \bm{r}_{i} \rbrace$ to $\lbrace \bm{q}_{i} \rbrace$: $- \partial \mathcal{U}(\lbrace \bm{r}_{i} \rbrace) / \partial \bm{r}_{i} = - \partial \mathcal{U}(\lbrace \bm{q}_{i} \rbrace) / \partial \bm{q}_{i}$.
From eqs \eqref{langevin_equation}, \eqref{langevin_equation_cm}, and \eqref{relative_position},
the dynamic equation for the relative position becomes 
\begin{equation}
 \label{langevin_equation_relative}
 \frac{d\bm{q}_{i}(t)}{dt} = - \frac{1}{\zeta} \frac{\partial  \mathcal{U}(\lbrace \bm{q}_{i}(t) \rbrace)}{\partial \bm{q}_{i}(t)}
  + \bm{\kappa} \cdot \bm{q}_{i}(t) 
  + \sqrt{\frac{2 k_{B} T}{\zeta}} \bm{W}_{i}(t),
\end{equation}
where $\bm{W}_{i}(t) = \bm{w}_{i}(t) - \bar{\bm{w}}(t)$ is again
a Gaussian white noise.
The first and second order moments of $\bm{W}_{i}(t)$ are calculated as follows:
\begin{align}
 \langle \bm{W}_{i}(t) \rangle & = \langle \bm{w}_{i}(t) \rangle - \langle \bar{\bm{w}}(t) \rangle = 0, \\
\begin{split}
 \label{fdr_second_order_relative_position}
  \langle \bm{W}_{i}(t)\bm{W}_{j}(t') \rangle 
 & = \langle \bm{w}_{i}(t) \bm{w}_{j}(t') \rangle 
 - \langle \bm{w}_{i}(t) \bar{\bm{w}}(t) \rangle \\
 & \qquad - \langle \bar{\bm{w}}(t) \bm{w}_{j}(t) \rangle 
 + \langle \bar{\bm{w}}(t) \bar{\bm{w}}(t') \rangle  \\
 & = \left(\delta_{ij} - \frac{1}{N}\right) \bm{1} \delta(t - t').
\end{split} 
\end{align}
The second order moment of the noises for different beads are coupled.
This is due to the fact that $\bm{q}_{i}$ is not linearly independent.

Eq~\eqref{langevin_equation} can be replaced by eqs~\eqref{langevin_equation_cm} and
\eqref{langevin_equation_relative}, and the contribution of the center of mass position to
the stress tensor is trivial (eq~\eqref{stress_tensor_cm}).
Thus we need only eq~\eqref{langevin_equation_relative} to analyze the
stress tensor of the Rouse-type model.
The stress tensor of the system can be decomposed into two contribuitions: the
stress by the center of mass and that by the relative positions.
By subtracting eq~\eqref{stress_tensor_cm} from eq~\eqref{stress_tensor},
we have the latter as
\begin{equation}
 \label{stress_tensor_relative_position}
 \bm{\sigma} - \bm{\sigma}_{\text{eq}} 
  = \frac{\rho}{N}
  \left[ \sum_{i = 1}^{N} \frac{\partial  \mathcal{U}(\lbrace \bm{q}_{i} \rbrace)}{\partial \bm{q}_{i}} \bm{q}_{i}
  - (N - 1) k_{B} T \bm{1} \right].
\end{equation}
Here, we have utilized the relation $\sum_{i = 1}^{N} \partial \mathcal{U}(\lbrace \bm{q}_{i} \rbrace) / \partial \bm{q}_{i} = 0$.

\subsection{Time Derivative of Gyration Tensor}

The gyration tensor \eqref{gyration_tensor} can be rewritten
in terms of the relative positions as
\begin{equation}
 \label{gyration_tensor_relative_position}
  \bm{S} = \frac{1}{N} \sum_{i = 1}^{N} \langle \bm{q}_{i} \bm{q}_{i} \rangle.
\end{equation}
From eq~\eqref{gyration_tensor_relative_position}, it is clear that
we can easily calculate the gyration tensor if we have the explicit expression
of $\langle \bm{q}_{i} \bm{q}_{i} \rangle$.
Thus we consider the time derivative of a diad $\bm{q}_{i}(t) \bm{q}_{i}(t)$.
From the Ito formula\cite{Gardiner-book} and eq~\eqref{fdr_second_order_relative_position}, we have
\begin{equation}
 \label{time_derivative_qi_qi}
  \begin{split}
  \frac{d}{dt} [\bm{q}_{i}(t) \bm{q}_{i}(t)] 
 & = \frac{d \bm{q}_{i}(t)}{dt} \bm{q}_{i}(t)
   + \bm{q}_{i}(t) \frac{d \bm{q}_{i}(t)}{dt}  \\
 & \qquad + \frac{1}{2} 2 \times \frac{2 k_{B} T}{\zeta} 
   \left(1 - \frac{1}{N}\right) \bm{1} .
 \end{split}
\end{equation}
The statistical average of \eqref{time_derivative_qi_qi} becomes
\begin{equation}
 \label{time_derivative_qi_qi_average}
  \begin{split}
   & \frac{d}{dt} \langle \bm{q}_{i}(t) \bm{q}_{i}(t) \rangle \\
 & = - \frac{1}{\zeta} \left\langle \frac{\partial \mathcal{U}(\lbrace \bm{q}_{i}(t) \rbrace)}{\partial  \bm{q}_{i}(t)} \bm{q}_{i}(t) 
   - k_{B} T \frac{N - 1}{N} \bm{1} \right\rangle \\
 & \qquad - \frac{1}{\zeta} \left\langle \bm{q}_{i}(t) \frac{\partial \mathcal{U}(\lbrace \bm{q}_{i}(t) \rbrace)}{\partial  \bm{q}_{i}(t)} 
   - k_{B} T \frac{N - 1}{N} \bm{1} \right\rangle \\
 & \qquad    + \bm{\kappa} \cdot \left\langle \bm{q}_{i}(t)\bm{q}_{i}(t) \right\rangle 
   + \left\langle \bm{q}_{i}(t)\bm{q}_{i}(t) \right\rangle \cdot  \bm{\kappa}^{\mathrm{T}}.
 \end{split}
\end{equation}

By taking the summation of eq~\eqref{time_derivative_qi_qi_average} over $i$
and using the expression of the stress tensor by eq~\eqref{stress_tensor_relative_position},
we have the following simple relation:
\begin{equation}
 \label{time_dependent_general_relation_stress_gyration}
 \begin{split}
  \frac{d \bm{S}(t)}{dt} 
  & = - \frac{2}{\rho \zeta} [\bm{\sigma}(t) - \bm{\sigma}_{\text{eq}}]
  + \bm{\kappa} \cdot \bm{S}(t)
  + \bm{S}(t) \cdot \bm{\kappa}^{\mathrm{T}}.
 \end{split}
\end{equation}
By rearranging terms, eq~\eqref{time_dependent_general_relation_stress_gyration} can be rewritten as follows:
\begin{equation}
 \label{time_dependent_general_relation_stress_gyration_inverted}
  \overset{\nabla}{\bm{S}}(t) =  - \frac{2} {\rho \zeta} [\bm{\sigma}(t) - \bm{\sigma}_{\text{eq}}] ,
\end{equation}
with the upper-convected time derivative defined as
\begin{equation}
  \overset{\nabla}{\bm{S}}(t) = \frac{d \bm{S}(t)}{dt} - \bm{\kappa} \cdot \bm{S}(t)
  - \bm{S}(t) \cdot \bm{\kappa}^{\mathrm{T}} .
\end{equation}
Eq~\eqref{time_dependent_general_relation_stress_gyration_inverted}
resembles to the upper-convected Maxwell model, which is one of the
simplest constitutive equation models\cite{Ottinger-book}.
However, unlike the upper-convected Maxwell model, we have no
explicit relation between $\bm{S}(t)$ and $\bm{\sigma}(t)$ in the Rouse-type model.
Thus eq~\eqref{time_dependent_general_relation_stress_gyration_inverted}
cannot be used as a closed constitutive equation.
It should be interpreted rather as a relation which connects the gyration
tensor and the stress tensor.


At the steady state, $\bm{\sigma}$ and $\bm{S}$ become independent of $t$.
They can be interpreted as functions of $\bm{\kappa}$, and we express
the steady state values as $\bm{\sigma}(\bm{\kappa})$
and $\bm{S}(\bm{\kappa})$.
Then eq \eqref{time_dependent_general_relation_stress_gyration} reduces to
\begin{equation}
 \label{steady_general_relation_stress_gyration}
 \bm{\sigma}(\bm{\kappa}) - \bm{\sigma}_{\text{eq}} =  \frac{\rho \zeta}{2} 
  \left[ \bm{\kappa} \cdot \bm{S}(\bm{\kappa}) +
   \bm{S}(\bm{\kappa}) \cdot \bm{\kappa}^{\mathrm{T}} \right].
\end{equation}
This simple relation holds for any interaction potentials,
as long as the dynamic equation is given as eq~\eqref{langevin_equation}.
To the best of the author's knowledge, these simple yet general relations have
not been reported so far.

It would be fair to mention that similar relations have been reported
for some simple models such as the dumbbell model\cite{Bird-Curtis-Armstrong-Hassager-book}
and the Rouse-like model with harmonic springs\cite{Jiang-2024}.
We stress that our relations (eqs~\eqref{time_dependent_general_relation_stress_gyration},
\eqref{time_dependent_general_relation_stress_gyration_inverted},
and \eqref{steady_general_relation_stress_gyration}) can be applied
to more general systems.

\section{RESULTS}
\label{results}

We apply the general relations obtained in Sec.~\ref{analysis}
(eqs~\eqref{time_dependent_general_relation_stress_gyration_inverted}
and \eqref{steady_general_relation_stress_gyration}) to some simple cases.
We consider the steady shear and steady uniaxial elongational flows, and
also the start-up and shut-down shear flows.

\subsection{Steady Shear}

\begin{figure}[tb]
\begin{center}
 \includegraphics[width=0.849\refwidth]{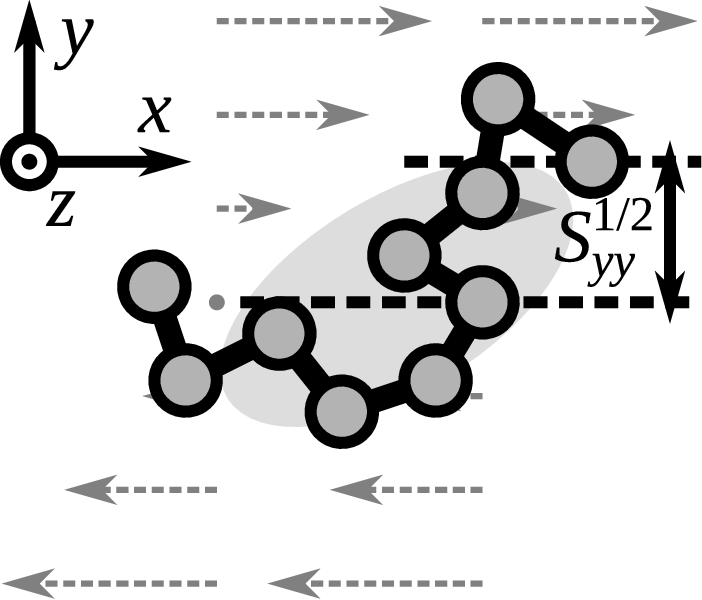}%
\end{center}
\caption{The Rouse-type model under a steady shear flow of which velocity
 gradient is given as eq~\eqref{velocity_gradient_tensor_shear}.
 Gray dashed arrows represent flow fields, and the gray ellipse represents the gyration tensor $\bm{S}$. 
 The flow and shear gradient directions are the $x$- and $y$-directions, respectively.
The steady shear viscosity $\eta(\dot{\gamma})$
 is governed by the $yy$-component of the gyration tensor, $S_{yy}(\dot{\gamma})$.
 \label{rouse_type_model_shear} (eq~\eqref{steady_shear_viscosity})}
\end{figure}

We consider the steady shear flow where the velocity gradient tensor is given as
\begin{equation}
 \label{velocity_gradient_tensor_shear}
 \kappa_{\alpha\beta} =
  \begin{cases}
   \dot{\gamma} & (\alpha = x, \beta = y), \\
   0 & (\text{otherwise}).
  \end{cases}  
\end{equation}
Here we have set the flow direction to the $x$-direction and
the shear gradient direction to the $y$-direction.
The images of the flow field and the gyration tensor are shown
in Figure~\ref{rouse_type_model_shear}.
In this case, eq~\eqref{steady_general_relation_stress_gyration}
can be further simplified as
\begin{align}
 \label{steady_general_relation_stress_gyration_shear_xy}
  \sigma_{xy}(\dot{\gamma}) & =  \frac{\rho \zeta}{2} \dot{\gamma} S_{yy}(\dot{\gamma}), \\
 \label{steady_general_relation_stress_gyration_shear_xx}
  \sigma_{xx}(\dot{\gamma}) & =  \rho \zeta \dot{\gamma} S_{xy}(\dot{\gamma}) - \frac{\rho}{N} k_{B} T, \\
 \label{steady_general_relation_stress_gyration_shear_yy}
  \sigma_{yy}(\dot{\gamma}) & = - \frac{\rho}{N} k_{B} T.
\end{align}
Eqs \eqref{steady_general_relation_stress_gyration_shear_xy}-\eqref{steady_general_relation_stress_gyration_shear_yy}
give the steady state viscosity and first normal stress coefficient:
\begin{align}
 \label{steady_shear_viscosity}
 \eta(\dot{\gamma}) &= \frac{\sigma_{xy}(\dot{\gamma})}{\dot{\gamma}} = \frac{\rho \zeta}{2} S_{yy}(\dot{\gamma})
 = \frac{\rho \zeta}{2} R_{g,y}^{2}(\dot{\gamma}), \\
 \label{steady_first_normal_stress_coefficient}
 \Psi_{1}(\dot{\gamma}) &= \frac{\sigma_{xx}(\dot{\gamma}) - \sigma_{yy}(\dot{\gamma})}{\dot{\gamma}^{2}} = \frac{\rho \zeta}{\dot{\gamma}} S_{xy}(\dot{\gamma}).
\end{align}

Eq~\eqref{steady_shear_viscosity} is one of the main results of this work.
It can be interpreted as a formula which gives the steady shear viscosity
in terms of the gyration tensor.
The steady shear viscosity $\eta(\dot{\gamma})$ is determined by $S_{yy}(\dot{\gamma}) = R_{g,y}^{2}(\dot{\gamma})$,
{\em not} by $S_{xy}(\dot{\gamma})$. 
This means that the shear viscosity is governed by the gyration radius {\em in the shear gradient
direction}. What is governed by $S_{xy}(\dot{\gamma})$ is the
steady first normal stress coefficient (eq~\eqref{steady_first_normal_stress_coefficient}).
These results seem not to be consistent with our naive expectation based on
the stress-optical rule\cite{Doi-Edwards-book}.
If we assume that the gyration tensor reflects the bond conformations,
we expect that the shear viscosity is related to $S_{xy}(\dot{\gamma}) / \dot{\gamma}$
and the first normal stress coefficient is related to $[S_{xx}(\dot{\gamma}) - S_{yy}(\dot{\gamma})] / \dot{\gamma}^{2}$, as
rough yet intuitive approximations.
We stress that we employed no approximations in the derivation of eqs~\eqref{steady_shear_viscosity}
and \eqref{steady_first_normal_stress_coefficient}.
Thus our results are more
general and physical reasonable than the relations naively expected from the stress-optical rule,
even if they are counter-intuitive.
Eqs~\eqref{steady_shear_viscosity} and \eqref{steady_first_normal_stress_coefficient} hold
for any interaction potentials, even if the stress-optical rule does not hold.

\subsection{Steady Uniaxial Elongation}

\begin{figure}[tb]
\begin{center}
 \includegraphics[width=0.748\refwidth]{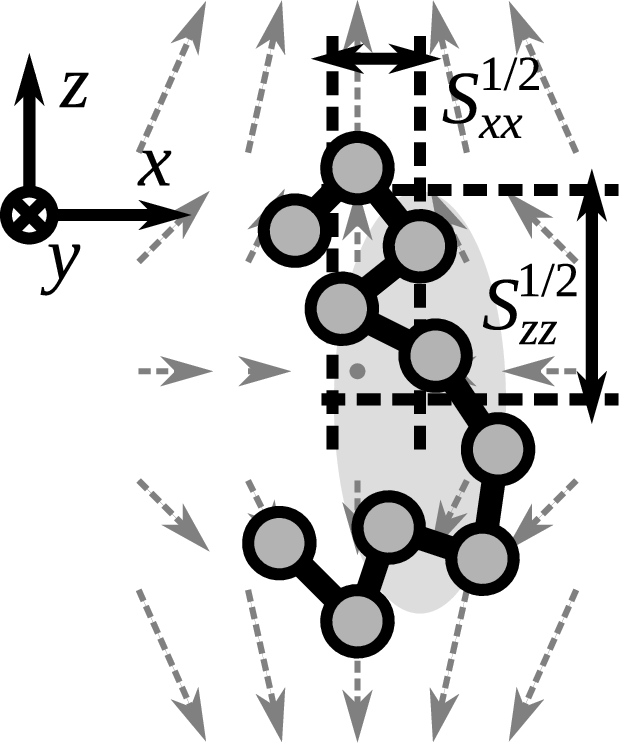}%
\end{center}
\caption{The Rouse-type model under a steady uniaxial elongation shear flow of which velocity
 gradient is given as eq~\eqref{velocity_gradient_tensor_uniaxial_elongation}.
 Gray dashed arrows represent flow fields, and the gray ellipse represents the gyration tensor $\bm{S}$. 
 The elongation direction is the $z$-direction.
 The steady elongational viscosity $\eta_{E}(\dot{\epsilon})$
 is governed by weighted sum of the $zz$- and $xx$-components of the gyration tensor, $S_{zz}(\dot{\epsilon}) + S_{xx}(\dot{\epsilon}) / 2$
 (eq~\eqref{steady_elongational_viscosity}).
 \label{rouse_type_model_elongation}}
\end{figure}

We consider the steady uniaxial elongation flow. We set the velocity gradient
tensor as
\begin{equation}
 \label{velocity_gradient_tensor_uniaxial_elongation}
 \kappa_{\alpha\beta}
  = \begin{cases}
     - \dot{\epsilon} / 2 & (\alpha = \beta = x, y), \\
     \dot{\epsilon} & (\alpha = \beta = z), \\
     0 & (\text{otherwise}).
    \end{cases}
\end{equation}
The elongation direction is set to the $z$-direction.
The images of the flow field and the gyration tensor are shown
in Figure~\ref{rouse_type_model_elongation}.
In this case, eq~\eqref{steady_general_relation_stress_gyration}
gives
\begin{align}
 \label{steady_general_relation_stress_gyration_elongation_zz}
 \sigma_{zz}(\dot{\epsilon}) & =  \rho \zeta \dot{\epsilon} S_{zz}(\dot{\epsilon}) - \frac{\rho}{N} k_{B} T, \\
 \label{steady_general_relation_stress_gyration_elongation_xx}
 \sigma_{xx}(\dot{\epsilon})
 & = \sigma_{yy}(\dot{\epsilon})
 = - \frac{\rho \zeta}{2} \dot{\epsilon} S_{xx}(\dot{\epsilon})
 - \frac{\rho}{N} k_{B} T.
\end{align}
We have utilized the relation $S_{xx}(\dot{\epsilon}) = S_{yy}(\dot{\epsilon})$.
From eqs~\eqref{steady_general_relation_stress_gyration_elongation_zz}
and \eqref{steady_general_relation_stress_gyration_elongation_xx}, we have
the following formula for the steady elongational viscosity:
\begin{equation}
\begin{split}
 \label{steady_elongational_viscosity}
 \eta_{E}(\dot{\epsilon}) 
 &  = \frac{\sigma_{zz}(\dot{\epsilon}) - \sigma_{xx}(\dot{\epsilon})}{\dot{\epsilon}}
  = \rho \zeta 
  \left[ S_{zz}(\dot{\epsilon}) + \frac{1}{2} S_{xx}(\dot{\epsilon}) \right] \\
 & = \rho \zeta
  \left[ R_{g,z}^{2}(\dot{\epsilon}) + \frac{1}{2} R_{g,x}^{2}(\dot{\epsilon}) \right].
\end{split}
\end{equation}

Eq~\eqref{steady_elongational_viscosity} can be interpreted as the formula
for the steady elongational viscosity.
It means that the steady
elongational viscosity is governed by the {\em sum} of 
$R_{g,z}^{2}(\dot{\epsilon})$ and $R_{g,x}^{2}(\dot{\epsilon}) / 2$,
not by the {\em difference} of them.
Although this result may be counter-intuitive again,
it is more general and physical reasonable than the naive expectation based on the stress-optical rule.

\subsection{Steady Biaxial and Planar Elongations}

In addition to the shear and uniaxial elongation flows,
biaxial and planar elongation flows are used in some MD simulations.
Here we consider the steady biaxial and planar elongational viscosties.
The calculation can be done in a similar way to the case of the
uniaxial elongational flow.

For the biaxial elongational flow, we set the velocity gradient
tensor as
\begin{equation}
 \label{velocity_gradient_tensor_biaxial_elongation}
 \kappa_{\alpha\beta}
  = \begin{cases}
     \dot{\epsilon}_{B} & (\alpha = \beta = x, y), \\
     - 2 \dot{\epsilon}_{B} & (\alpha = \beta = z), \\
     0 & (\text{otherwise}),
    \end{cases}
\end{equation}
where $\dot{\epsilon}_{B}$ is the biaxial strain rate.
Eq~\eqref{velocity_gradient_tensor_biaxial_elongation} corresponds to
eq~\eqref{velocity_gradient_tensor_uniaxial_elongation} with
$\dot{\epsilon}_{B} = - \dot{\epsilon} / 2$. The stress tensor can
be calculated in the same way as
eqs~\eqref{steady_general_relation_stress_gyration_elongation_zz}-\eqref{steady_general_relation_stress_gyration_elongation_xx}.
The steady biaxial elongational viscosity becomes
\begin{equation}
\begin{split}
 \label{steady_biaxial_elongational_viscosity}
 \eta_{B}(\dot{\epsilon}_{B}) 
 &  = \frac{\sigma_{xx}(\dot{\epsilon}_{B}) - \sigma_{zz}(\dot{\epsilon}_{B})}{\dot{\epsilon}_{B}} \\
 & = \rho \zeta 
  \left[ S_{xx}(\dot{\epsilon}_{B}) + 2 S_{zz}(\dot{\epsilon}_{B}) \right] \\
 & = \rho \zeta
  \left[ R_{g,x}^{2}(\dot{\epsilon}_{B}) + 2 R_{g,z}^{2}(\dot{\epsilon}_{B}) \right].
\end{split}
\end{equation}

For the planar elongational flow, we set the velocity gradient tensor
as
\begin{equation}
 \label{velocity_gradient_tensor_planar_elongation}
 \kappa_{\alpha\beta}
  = \begin{cases}
     \dot{\epsilon}_{P} & (\alpha = \beta = x), \\
     - \dot{\epsilon}_{P} & (\alpha = \beta = y), \\
     0 & (\text{otherwise}),
    \end{cases}
\end{equation}
where $\dot{\epsilon}_{P}$ is the planar strain rate. The steady
planar elongational viscosity can be calculated in a similar way.
The result is
\begin{equation}
\begin{split}
 \label{steady_palanar_elongational_viscosity}
 \eta_{P}(\dot{\epsilon}_{P}) 
 &  = \frac{\sigma_{xx}(\dot{\epsilon}_{P}) - \sigma_{yy}(\dot{\epsilon}_{P})}{\dot{\epsilon}_{P}} \\
 & = \rho \zeta 
  \left[ S_{xx}(\dot{\epsilon}_{P}) + S_{yy}(\dot{\epsilon}_{P}) \right] \\
 & = \rho \zeta
  \left[ R_{g,x}^{2}(\dot{\epsilon}_{P}) + R_{g,y}^{2}(\dot{\epsilon}_{P}) \right].
\end{split}
\end{equation}

From eqs~\eqref{steady_biaxial_elongational_viscosity} and
\eqref{steady_palanar_elongational_viscosity}, we find that the steady
viscosities depend on the weighted sum of the gyration radii in different
directions. This situation is similar to that of the steady elongational
viscosity (eq~\eqref{steady_elongational_viscosity}).
However, the weight factors are not common. They depend on the flow type.

\subsection{Start-Up Shear and Shut-Down Shear}

By using eq~\eqref{time_dependent_general_relation_stress_gyration_inverted},
we can calculate the expression for the viscosity growth function
$\eta^{+}(t;\dot{\gamma}) = {\sigma^{+}_{xy}(t;\dot{\gamma})} / {\dot{\gamma}}$
at the start-up process.
We assume that the system is in equilibrium at $t = 0$, and
set the velocity gradient tensor $\bm{\kappa}$ as eq~\eqref{velocity_gradient_tensor_shear}
for $t > 0$.
The viscosity growth function is expressed in terms of the 
gyration tensor growth function $\bm{S}^{+}(t;\dot{\gamma})$ as
\begin{equation}
 \label{viscosity_growth_function}
  \eta^{+}(t;\dot{\gamma})
  = \frac{\rho \zeta}{2} 
  \left[ S_{yy}^{+}(t;\dot{\gamma}) - \frac{1}{\dot{\gamma}} \frac{d S_{xy}^{+}(t;\dot{\gamma})}{dt}  \right].
\end{equation}
It is straightforward to show that eq~\eqref{viscosity_growth_function}
reduces to eq~\eqref{steady_shear_viscosity} at the limit of $t \to \infty$.
Eq \eqref{viscosity_growth_function}
can be rewritten as
\begin{equation}
 \label{time_dependent_general_relation_stress_gyration_shear_xy_inverted}
  \frac{d S_{xy}^{+}(t;\dot{\gamma})}{dt} 
  = \frac{2\dot{\gamma}}{\rho \zeta} \left[ \frac{\rho \zeta}{2} S^{+}_{yy}(t;\dot{\gamma}) - \eta^{+}(t;\dot{\gamma}) \right].
\end{equation}
Eq~\eqref{time_dependent_general_relation_stress_gyration_shear_xy_inverted}
means that the time-evolution of $S_{xy}^{+}(t;\dot{\gamma})$
is determined by the degree of violation of
eq~\eqref{steady_shear_viscosity} at the transient state.

In a similar way, we can calculate the first normal stress growth function 
$\Psi_{1}^{+}(t;\dot{\gamma}) = [\sigma_{xx}^{+}(t;\dot{\gamma}) - \sigma_{yy}^{+}(t;\dot{\gamma})] / \dot{\gamma}^{2}$
as
\begin{equation}
 \label{first_normal_stress_coefficient_growth_function}
  \Psi^{+}_{1}(t;\dot{\gamma}) 
  =  \frac{\rho \zeta}{\dot{\gamma}}
 \left[ 
 S_{xy}^{+}(t;\dot{\gamma}) 
 - \frac{1}{2 \dot{\gamma}} \frac{d[S_{xx}^{+}(t;\dot{\gamma}) - S_{yy}^{+}(t;\dot{\gamma})] }{dt} 
 \right].
\end{equation}
Of course, eq~\eqref{first_normal_stress_coefficient_growth_function}
reduces to \eqref{steady_first_normal_stress_coefficient} at the limit of $t \to \infty$.
As before,
eqs~\eqref{viscosity_growth_function} and \eqref{first_normal_stress_coefficient_growth_function}
holds for any interaction potentials.
They will be useful to calculate time-dependent rheological quantities based on the gyration tensor
growth function.

The shut-down process can be handled in a similar way.
We assume that the system is in the steady state with
the velocity gradient tensor given by eq~\eqref{velocity_gradient_tensor_shear}
at $t = 0$.
Then we set $\bm{\kappa} = 0$ for $t > 0$.
From eq~\eqref{time_dependent_general_relation_stress_gyration_inverted},
we have the following simple relation between the stress decay function
$\bm{\sigma}^{-}(t;\dot{\gamma})$ and the gyration tensor decay function
$\bm{S}^{-}(t;\dot{\gamma})$:
\begin{equation}
 \frac{d\bm{S}^{-}(t;\dot{\gamma})}{dt} = - \frac{2}{\rho \zeta} \bm{\sigma}^{-}(t;\dot{\gamma}),
\end{equation}
Then the viscosity and first normal stress coefficient decay functions become
\begin{align}
 \label{viscosity_decay_function_shear}
 \eta^{-}(t;\dot{\gamma}) & = -  \frac{\rho \zeta}{2 \dot{\gamma}} \frac{dS_{xy}^{-}(t;\dot{\gamma})}{dt}, \\
 \Psi_{1}^{-}(t;\dot{\gamma}) & = - \frac{\rho \zeta}{2 \dot{\gamma}^{2}} \frac{d[S_{xx}^{-}(t;\dot{\gamma}) - S_{yy}^{-}(t;\dot{\gamma})]}{dt}.
\end{align}
%
%

\section{DISCUSSIONS}
\label{discussions}

\subsection{Zero Strain Rate Limit}

At the limit of $\bm{\kappa} \to 0$, the gyration tensor can be
approximated by the equilibrium value: $\bm{S}(\bm{\kappa} \to 0) = (R_{g,\text{eq}}^{2} / 3) \bm{1}$
($R_{g,\text{eq}}$ is the gyration radius in equilibrium).
In this zero strain rate limit, the steady shear viscosity reduces
to the zero shear viscosity: $\eta_{0} = \eta(\dot{\gamma} \to 0)$.
The steady elongational viscosity reduces to $\eta_{E}(\dot{\epsilon} \to 0) = 3 \eta_{0}$
(so-called the Trouton's rule).

From eq~\eqref{steady_shear_viscosity}, we have the following simple
expression for the zero shear viscosity:
\begin{equation}
 \label{zero_shear_viscosity}
 \eta_{0} = \frac{\rho \zeta}{6} R_{g,\text{eq}}^{2}.
\end{equation}
In a similar way, from eq~\eqref{steady_elongational_viscosity}, we have
\begin{equation}
 \label{zero_rate_steady_elongational_viscosity}
 \eta_{E}(\dot{\epsilon} \to 0) = \frac{\rho \zeta}{2} R_{g,\text{eq}}^{2} = 3 \eta_{0}.
\end{equation}
Thus the Trouton's rule is correctly reproduced.

Eq~\eqref{zero_shear_viscosity} means that the zero shear viscosity
is governed by $R_{g,\text{eq}}$.
A relation similar to eq~\eqref{zero_shear_viscosity} was already
proposed by Xu, Chen, and An\cite{Xu-Chen-An-2015} based on MD simulation data
for unentangled polymers with various architectures: $\eta_{0} \propto R_{g,\text{eq}}^{2}$.
(They cound not identify the prefactor $\rho \zeta / 2$, however.)
If we have a series of unentangled
polymers with the same monomers and different architectures, polymers
with small $R_{g,\text{eq}}$ exhibit low viscosities.
For example, a star polymer exhibits lower viscosity than a linear polymer
with the same total molecular weight.

The growth functions can be simplified at the limit of $\bm{\kappa} \to 0$.
The stress growth function $\eta^{+}(t;\dot{\gamma})$ can be related to the
shear relaxation modulus $G(t)$ as $\eta^{+}(t;\dot{\gamma} \to 0) = \int_{0}^{t} dt' \, G'(t')$.
$S_{yy}^{+}(t,\dot{\gamma})$ can be also related to the shear relaxation modulus:
$S_{yy}^{+}(t;\dot{\gamma} \to 0) = R_{g,\text{eq}}^{2} / 3 = 2 \eta_{0} / \rho \zeta
= (2 / \rho \zeta) \int_{0}^{\infty} dt \, G(t)$.
Then, from eqs~\eqref{time_dependent_general_relation_stress_gyration_shear_xy_inverted}
and \eqref{zero_shear_viscosity},
we have
\begin{equation}
 \label{zero_shear_gyration_tensor_growth_xy}
  \begin{split}
   \lim_{\dot{\gamma} \to 0} \frac{S^{+}_{xy}(t;\dot{\gamma})}{ \dot{\gamma}} 
  & = \frac{2}{\rho \zeta} 
    \int_{0}^{t} dt'\left[ \eta_{0} - \int_{0}^{t'} dt'' \, G(t'') 
   \right] \\
   &  = \frac{2}{\rho \zeta} 
    \int_{0}^{t} dt' \int_{t'}^{\infty} dt'' \, G(t'') .
  \end{split}
\end{equation}
At the limit of $t \to \infty$, the growth function approaches to the steady value.
The steady state value $S_{xy}(\dot{\gamma})$ becomes
\begin{equation}
 \label{zero_shear_steady_gyration_tensor_xy}
  \begin{split}
 \lim_{\dot{\gamma} \to 0} \frac{S_{xy}(\dot{\gamma})}{\dot{\gamma}} 
   & = \frac{2}{\rho \zeta} 
    \int_{0}^{\infty} dt' \int_{t'}^{\infty} dt'' \, G(t'') \\
   & = \frac{2}{\rho \zeta} 
    \int_{0}^{\infty} dt \, t  G(t).
  \end{split}
\end{equation}
Eqs~\eqref{zero_shear_gyration_tensor_growth_xy} and
\eqref{zero_shear_steady_gyration_tensor_xy}
give $S^{+}_{xy}(t;\dot{\gamma})$ and $S_{xy}(\dot{\gamma})$
in terms of $G(t)$ for sufficiently small $\dot{\gamma}$.

\subsection{Rouse Model with Harmonic Springs}

Here we apply our formula \eqref{steady_shear_viscosity} to
the Rouse model with harmonic springs. We set to interaction potential
as
\begin{equation}
 \mathcal{U}(\lbrace \bm{q}_{i} \rbrace) = \sum_{i = 1}^{N - 1} \frac{3 k_{B} T}{2 b^{2}} (\bm{q}_{i + 1} - \bm{q}_{i})^{2},
\end{equation}
where $b$ is the segment size.
The steady shear viscosity
of the Rouse model is known to be independent of the shear rate.
The shear viscosity is simply the same as the zero-shear viscosity $\eta_{0}$,
and it is given as
\begin{equation}
 \label{steady_shear_viscosity_rouse_model}
 \eta(\dot{\gamma}) = \eta_{0} = \frac{\rho (N - 1) \zeta b^{2}}{36}.
\end{equation}

We show that our formula successfully reproduces eq~\eqref{steady_shear_viscosity_rouse_model}.
Due to the linear nature of the Rouse model, the chain dimension in
the $y$-direction is not changed under a shear flow. Then, the gyration
radius in the $y$-direction is simply calculated from the equilibrium gyration radius:
\begin{equation}
 \label{gyration_radius_y_rouse_model}
 R_{g,y}^{2}(\dot{\gamma}) = \frac{1}{3} R_{g,\text{eq}}^{2}
 = \frac{1}{18} (N - 1) b^{2}.
\end{equation}
where we have utilized $R_{g,\text{eq}}^{2} = (N - 1) b^{2} / 6$.
It is now straightforward that our formula \eqref{steady_shear_viscosity}
combined with eq~\eqref{gyration_radius_y_rouse_model}
gives eq~\eqref{steady_shear_viscosity_rouse_model}.

\subsection{Comparison with Molecular Dynamics Simulation Data in Literature}

MD simulations are useful to study the conformational
and rheological behaviors of unentangled polymers. We can find
various simulation data in the literature. Here we briefly discuss some
literature data from the viewpoint of our formula.

Xu, Chen, and An\cite{Xu-Chen-An-2017} studied the relation
between the steady shear viscosities and chain conformations
of unentangled polymers with various architectures. They reported that
the steady shear viscosity is proportional to $S_{yy}(\dot{\gamma})$ as
$\eta(\dot{\gamma}) \propto S_{yy}^{3/2}(\dot{\gamma})$. Although
the exponent $3/2$ is different from our formula \eqref{steady_shear_viscosity}
(which gives the exponent $1$), it is common
that $S_{yy}(\dot{\gamma})$ governs the shear viscosity.
(The discrepancy between their empirical relation and our theoretical
prediction may be attributed to the friction reduction effect. We will
discuss it in Sec.~5.4.)
G\"{u}rel and Giuntoli\cite{Gurel-Giuntoli-2023} analyzed the
correlation between various quantities under shear and reported
that the steady shear viscosity is strongly correlated to $S_{yy}(\dot{\gamma})$.
However, they interpreted that this correlation is a secondary effect.
Based on our formula \eqref{steady_shear_viscosity}, it is not a secondary
effect but nothing but the primary effect.
Recently, Oishi and coworkers\cite{Oishi-Koide-Ishida-Uneyama-Masubuchi-MullerPlathe-2024} performed reverse non-equilibrium
MD (RNEMD) simulations to study the steady shear viscosity
of an unentangled linear polymer. They calculated both the steady shear viscosity
data and the gyration radius data. The steady shear viscosity data
seem to be well correlated to the gyration radius in the $y$-direction
(compare Figures~5 and 10 of Ref.~\cite{Oishi-Koide-Ishida-Uneyama-Masubuchi-MullerPlathe-2024}).
From these literature data, we consider that our formulae are
reasonable.

In MD simulations, gyration tensor data can be accurately obtained from
just a single snapshot because we can take the average over many chains.
On the other hand, it is not easy to obtain accurate stress tensor data.
We have just one stress tensor from a single snapshot, and we need to
take an average over many snapshots at the steady state.
By using our formulae, we can estimate the viscosity data from
the gyration tensor data. Although our formulae based on the
Rouse-type model may not be fully applicable in some cases
(for example, where the friction reduction occurs, which will be discussed in the
next subsection), they will
be useful to roughly estimate viscosities under fast flows.

\subsection{Friction Reduction}

Recently, the reduction of the friction coefficient (or the
increase of the mobility) under fast flows is widely studied.
Although such a modulation is first proposed for
entangled polymers\cite{Uneyama-Horio-Watanabe-2011,Yaoita-Isaki-Masubuchi-Watanbe-Ianniruberto-Marrucci-2012},
later it is also applied to unentangled polymers\cite{Watanabe-Matsumiya-Sato-2021,Sato-Matsumiya-Watanabe-2022,Matsumiya-Sato-Chen-Watanabe-2022,Jiang-vanRuymbeke-2023}.
Here we discuss the generalization of our model to include the
friction reduction effect.

In the case where the friction coefficient is isotropically modulated
under a fast flow, we can replace $\zeta$ in eq~\eqref{langevin_equation}
by the effective friction coefficient under a fast flow.
In general, the fluctuation-dissipation relation can be violated under
a fast flow. Thus we introduce two effective friction coefficients
$\zeta_{\text{eff}}(\bm{\kappa})$ and $\zeta_{\text{eff}}'(\bm{\kappa})$
and modify eq~\eqref{langevin_equation} as
\begin{equation}
 \label{langevin_equation_with_friction_reduction}
  \begin{split}
  \frac{d\bm{r}_{i}(t)}{dt} & = - \frac{1}{\zeta_{\text{eff}}(\bm{\kappa})} \frac{\partial  \mathcal{U}(\lbrace \bm{r}_{i}(t) \rbrace)}{\partial \bm{r}_{i}(t)}
  + \bm{\kappa} \cdot \bm{r}_{i}(t) \\
   & \qquad + \sqrt{\frac{2 k_{B} T}{\zeta'_{\text{eff}}(\bm{\kappa})}} \bm{w}_{i}(t),
  \end{split}
\end{equation}
If the fluctuation-dissipation relation holds, we can set $\zeta_{\text{eff}}(\bm{\kappa}) = \zeta'_{\text{eff}}(\bm{\kappa})$.
The steady shear and elongational viscosities for eq~\eqref{langevin_equation_with_friction_reduction}
become
\begin{align}
 \label{steady_shear_viscosity_with_friction_reduction}
 \eta(\dot{\gamma}) & = \frac{\rho \zeta_{\text{eff}}(\dot{\gamma})}{2} R_{g,y}^{2}(\dot{\gamma}), \\
 \label{steady_elongational_viscosity_with_friction_reduction}
 \eta_{E}(\dot{\epsilon}) & = \rho \zeta_{\text{eff}}(\dot{\epsilon}) 
 \left[ R_{g,z}^{2}(\dot{\epsilon}) + \frac{1}{2} R_{g,x}^{2}(\dot{\epsilon}) \right].
\end{align}
Eqs~\eqref{steady_shear_viscosity_with_friction_reduction} and
\eqref{steady_elongational_viscosity_with_friction_reduction} are independent of $\zeta'_{\text{eff}}(\bm{\kappa})$,
and thus they are valid even if the fluctuation-dissipation relation is violated ($\zeta_{\text{eff}}(\bm{\kappa}) \neq \zeta'_{\text{eff}}(\bm{\kappa})$).

From eqs~\eqref{steady_shear_viscosity_with_friction_reduction} and
\eqref{steady_elongational_viscosity_with_friction_reduction},
we expect that the effective friction coefficients can be calculated
from the viscosity data and gyration radius data under flows.
The ratios of the effective friction coefficients to the equilibrium
friction coefficient become
\begin{align}
 \label{effective_friction_coefficient_steady_shear}
 \frac{\zeta_{\text{eff}}(\dot{\gamma})}{\zeta} & = \frac{\eta(\dot{\gamma})}{\eta_{0}} \frac{R_{g,\text{eq}}^{2}}{3 R_{g,y}^{2}(\dot{\gamma})}, \\
 \label{effective_friction_coefficient_steady_elongation}
  \frac{\zeta_{\text{eff}}(\dot{\epsilon})}{\zeta} & = 
  \frac{\eta_{E}(\dot{\epsilon})}{3 \eta_{0}}
 \frac{R_{g,\text{eq}}^{2}}{2 R_{g,z}^{2}(\dot{\epsilon}) + R_{g,x}^{2}(\dot{\epsilon})}.
\end{align}
Eqs~\eqref{effective_friction_coefficient_steady_shear} and
\eqref{effective_friction_coefficient_steady_elongation} can be used to
estimate the effective friction coefficients from MD simulation data.
For example, if we employ the empirical relation $\eta(\dot{\gamma}) \propto R_{g,y}^{3}(\dot{\gamma})$
by Xu, Chen, and An\cite{Xu-Chen-An-2017}, the effective friction coefficient 
obeys $\zeta_{\text{eff}}(\dot{\gamma}) \propto R_{g,y}(\dot{\gamma})
\propto \eta^{1/3}(\dot{\gamma})$.
An unentangled polymer melt is known to exhibits the shear thinning behavior,
and then the effective friction coefficient reduces under a fast shear.

In general, the friction coefficient becomes anisotropic under a fast flow\cite{Uneyama-Horio-Watanabe-2011,Watanabe-Matsumiya-Sato-2021}.
Thus we may employ the following form instead of eq~\eqref{langevin_equation_with_friction_reduction}
as the modified dynamic equation:
\begin{equation}
 \label{langevin_equation_with_friction_reduction_anisotropic}
\begin{split}
  \frac{d\bm{r}_{i}(t)}{dt} 
 & = - \bm{\Lambda}(\bm{\kappa}) \cdot \frac{\partial  \mathcal{U}(\lbrace \bm{r}_{i}(t) \rbrace)}{\partial \bm{r}_{i}(t)}
  + \bm{\kappa} \cdot \bm{r}_{i}(t) \\
 & \qquad + \sqrt{2 k_{B} T}
  \bm{B}(\bm{\kappa}) \cdot \bm{w}_{i}(t).
\end{split}
\end{equation}
Here, $\bm{\Lambda}(\bm{\kappa})$ is the effective mobility tensor which corresponds
to the inverse of the effective friction coefficient tensor, and
$\bm{B}(\bm{\kappa})$ is the effective noise coefficient tensor.
It would be convenient to introduce another effective mobility tensor
$\bm{\Lambda}'(\bm{\kappa}) = \bm{B}(\bm{\kappa}) \cdot   \bm{B}^{\mathrm{T}}(\bm{\kappa})$.
If the fluctuation-dissipation relation holds, we have
$\bm{\Lambda}(\bm{\kappa}) = \bm{\Lambda}'(\bm{\kappa})$.
Eq~\eqref{langevin_equation_with_friction_reduction} corresponds to the special
case of eq~\eqref{langevin_equation_with_friction_reduction_anisotropic}
with $\bm{\Lambda}(\bm{\kappa}) = \bm{1} / \zeta_{\text{eff}}(\bm{\kappa})$
and  $\bm{\Lambda}'(\bm{\kappa}) = \bm{1} / \zeta_{\text{eff}}'(\bm{\kappa})$.
For eq~\eqref{langevin_equation_with_friction_reduction_anisotropic},
the general relation between $\bm{S}(\bm{\kappa})$ and $\bm{\sigma}(\bm{\kappa})$ becomes as follows,
instead of eq~\eqref{steady_general_relation_stress_gyration}:
\begin{equation}
 \label{steady_general_relation_stress_gyration_friction_reduction_anisotropic}
  \begin{split}
   & \bm{\Lambda}(\bm{\kappa}) \cdot [\bm{\sigma}(\bm{\kappa}) - \bm{\sigma}_{\text{eq}}]
  + [\bm{\sigma}(\bm{\kappa}) - \bm{\sigma}_{\text{eq}}] \cdot \bm{\Lambda}(\bm{\kappa})  \\
   & =  \rho
  \left[ \bm{\kappa} \cdot \bm{S}(\bm{\kappa}) +
   \bm{S}(\bm{\kappa}) \cdot \bm{\kappa}^{\mathrm{T}} \right] \\
   & \qquad + 2 \rho k_{B} T \frac{N - 1}{N} [\bm{\Lambda}(\bm{\kappa})  - \bm{\Lambda}'(\bm{\kappa}) ].
  \end{split}
\end{equation}
In the left hand side of eq~\eqref{steady_general_relation_stress_gyration_friction_reduction_anisotropic},
different components of $\bm{\sigma}(\bm{\kappa})$ can be coupled
if $\bm{\Lambda}(\bm{\kappa})$ has nondiagonal components.
In addition, the violation of the fluctuation-dissipation relation ($\bm{\Lambda}(\bm{\kappa}) \neq \bm{\Lambda}'(\bm{\kappa})$) affects
the relation between $\bm{\sigma}(\bm{\kappa})$ and $\bm{S}(\bm{\kappa})$
via the last term in the right hand side of eq~\eqref{steady_general_relation_stress_gyration_friction_reduction_anisotropic}.
Thus the explicit expressions for the steady state viscosities will be complicated in this general case.
The detailed analysis of eq~\eqref{steady_general_relation_stress_gyration_friction_reduction_anisotropic} will
be interesting but beyond the scope of this work, and thus we do not go into further detail.

\section{CONCLUSIONS}
\label{conclusions}

We analyzed the rheological properties of the Rouse-type single chain model
for an unentangled polymer melt.
In the Rouse-type model, beads interact each other
via general nonlinear potentials such as the FENE and LJ potentials.
We used the relative position to the center of mass to derive the
general relations between the stress tensor $\bm{\sigma}$
and the gyration tensor $\bm{S}$
(eqs~\eqref{time_dependent_general_relation_stress_gyration_inverted}
and \eqref{steady_general_relation_stress_gyration}).
It is worth emphasized that we employed no additional approximations
to derive these relations.
Thus they can be used for various systems as long as the dynamic
equation is given as eq~\eqref{langevin_equation}.

We applied the general relations to some simple flows such as the
steady shear flow. The steady shear stress $\eta({\dot{\gamma}})$
was shown to be governed by the squared gyration radius in the shear
gradient direction, $R_{g,y}^{2}(\dot{\gamma})$.
Similar nontrivial relations were obtained for the first normal
stress coefficient, the steady elongational viscosity,
and the viscosity growth and decay functions.

We then discussed the the zero strain rate limit. At this limit,
we can calculate the growth function for the off-diagonal component
of the gyration tensor $S^{+}_{xy}(t;\dot{\gamma})$ only from
the shear modulus $G(t)$.
Well-known results such as the Trouton's rule and the shear viscosity
of the Rouse model can be successfully reproduced by our formulae.
In addition, our result seems to be qualitatively consistent with
the MD simulation data in the literature.

As a possible generalization, we discussed integration of the friction reduction
effect to the Rouse-type model. If we model the friction reduction as the isotropic modulation
of the friction coefficient, our result may be used to estimate the
effective friction coefficient under fast flows.
The analysis of the effective friction coefficient from
various MD simulation data will be an interesting future work.

\section*{ACKNOWLEDGMENT}
This work was supported by Grant-in-Aid (KAKENHI) for Scientific Research
Grant B No.~JP23H01142.


\end{document}